# Sustainability of largescale waste heat harvesting using thermoelectrics

Anilkumar Bohra and Satish Vitta*
Department of Metallurgical Engineering and Materials Science
Indian Institute of Technology Bombay
Mumbai 400076; India.

## Abstract

The amount of waste heat exergy generated globally is ~ 69.058 EJ which can be divided into, low temperature < 373 K, 30.496 EJ, medium temperature 373 K – 573 K, 14.431 EJ and high temperature > 573 K, 24.131 EJ. These values of exergy have been used to determine the minimum number of pn-junctions required to convert the exergy into electrical power. It is found that the number of junctions required to convert high temperature exergy increases from $8.22 \times 10^{11}$ to $24.66 \times 10^{11}$ when the aspect ratio of the legs increases from 0.5 $cm^{-1}$ to 1.5 $cm^{-1}$. To convert the low temperature exergy $81.76 \times 10^{11}$ to $245.25 \times 10^{11}$ junctions will be required depending on the legs aspect ratio. The quantity of alloys containing elements such as Pb, Bi, Te, Sb, Se and Sn required to synthesize these junctions therefore is of the order of 'millions of tons' which means the elements required is also of similar magnitude. The current world production of these elements however falls far short of this requirement, indicating significant supply chain risk. The production of these elements, even if resources are available, will emit millions of tons of $CO_2$ showing that current alloys are non-sustainable for waste heat recovery.

*Email: satish.vitta@iitb.ac.in

## 1) **Introduction**:

Energy utilization never takes place in it's primary form and it needs to be converted into usable forms. One of the most versatile forms of energy is 'electricity' which needs to be generated from a variety of primary resources such as coal, oil, gas, Sun, water, etc. Another form of energy which is used in large quantities is mechanical in nature which again requires to be generated from primary resources. Energy conversion processes in general have efficiencies ranging from ~ 20 % to 80 % with most of the processes used for large scale energy conversion having efficiency of the order of 30 % to 40 %.[1] This results in 60 % to 70 % of the energy in primary resources being lost and most of this energy loss is in the form of 'heat'.[2] Historically, many techniques have been developed to utilize at least part of this wasted heat.[3-5] One of the techniques which has been discussed often for conversion of waste heat into useful electricity is based on the Seebeck effect and is termed 'thermoelectric energy conversion'.[6-9] The significant advantage of this technique is that it has no moving parts which means no frictional loss of energy. The other significant merit of this technique is that once it is operationalized it can work for at least 30 years with waste heat as the only driving force and with sustainable conversion efficiency.[10,11] The limitation for large scale usage however has been attributed to 'non-availability' of materials with high 'figure-of-merit zT' and energy conversion efficiency $\eta$. The single most important factor that limits achieving a high zT has been the high thermal conductivity $\kappa$ as it is inversely proportional to zT. Recently, several studies, both experimental and theoretical, have been undertaken to understand $\kappa$ and hence design materials with low $\kappa$.[12] This has resulted in the development of several materials with a high figure of merit parameter $z_{max}$ at different temperatures as shown in Figure 1 and Tables S1-S3. These recent developments have brought this technology back into the forefront to harvest waste heat.

The objectives of the present work therefore have been twofold: 1) to assess its technological potential for large scale conversion of waste heat into electricity from the perspective of materials requirement and 2) to ascertain if zT is indeed the limiting factor for large scale implementation of waste heat recovery. The sustainability aspect has been assessed by investigating the embodied energy associated with the various elements required to manufacture the different alloys and the associated environmental emissions. The potential materials supply chain risk has also been analyzed by considering the current production capacity of different elements that are critical for the implementation of thermoelectric waste heat conversion and their Herfindahl-Hirschman Index.

**2.1) <u>Thermoelectric criteria</u>:**

The figure-of-merit zT of a thermoelectric material depends on the Seebeck coefficient $\alpha$, electrical conductivity $\sigma$ and thermal conductivity $\kappa$ and is given by the relation;

$$zT = \frac{\alpha^2 \sigma}{(\kappa_l + \kappa_e)} T \qquad (1)$$

where the thermal conductivity $\kappa$ is a combined contribution of phonons $\kappa_l$ and electrons $\kappa_e$. The technological potential of this energy conversion technology however depends on the power delivery capacity P of the device and its efficiency of energy conversion $\eta$. These are given by;

$$P = \frac{(\alpha_{pn} \Delta T)^2}{R_{TED}} \frac{s}{(1+s)^2} \qquad (2)$$

$$\eta = \frac{(\Delta T / T_h) s}{\left\{ (1+s) - \left( \Delta T / 2T_h \right) + \left[ (1+s)^2 / z_{pn} T_h \right] \right\}} \qquad (3)$$

where, $\alpha_{pn}$ and $z_{pn}$ are the effective Seebeck coefficient $|\alpha_p + \alpha_n|$ and figure-of-merit of the pn-junction device, $R_{TED}$ the total resistance of the device including the contacts, $\Delta T$ the temperature difference between hot and cold temperatures $T_h$ and $T_c$ and s the load resistance to device resistance ratio $R_L/R_{TED}$. It is seen that both power conversion as well as efficiency are strong functions of the resistance ratio s.[13] The power delivered by the device has a maximum for s = 1 while the efficiency peaks for $s = \left(1 + z_{pn} \bar{T}\right)^{1/2}$ showing that P and $\eta$ do not peak for the same conditions. The power delivered by the thermoelectric generator peaks when the load resistance $R_L$ becomes equal to that of the device $R_{TED}$, i.e. s = 1 and is given by;

$$P_{max}^N = \frac{(N \alpha_{pn} \Delta T)^2}{4 N R_{TED}} \qquad (4)$$

for a module consisting of N pn-junctions connected in series. The maximum power therefore depends only on the effective Seebeck coefficient $\alpha_{pn}$ and device resistance $R_{TED}$ given by;

$$R_{TED} = \left[\frac{\rho_p l_p}{A_p} + \frac{\rho_n l_n}{A_n} + \frac{2\rho_c l_c}{A_c}\right] \qquad (5)$$

where $\rho_p$, $\rho_n$ and $\rho_c$ are resistivities, $l_p$, $l_n$ and $l_c$ are leg lengths and $A_p$, $A_n$ and $A_c$ are cross-sectional areas of p-, n- and contact, respectively.

It has been proposed that effective power delivered by the thermoelectric device can be increased by reducing the device dimensionality, i.e. by nanostructuring. This results in enhancing the effective density-of-states for the same material, thus increasing the power factor. Absolute size of nanostructured features to enhance power factor beyond the bulk value is typically < 10 nm and depends on the material properties.[14] Several types of nanostructures such as artificial superlattices and nanowires have been shown to exhibit higher power factor concurrent with lowered thermal conductivity. This phenomenon however has not been observed universally and also processing the materials/devices into such structures requires complex processing steps which will add to energy costs of synthesis. Also, thermal stability of these structures will be a significant factor of technological concern. Hence, in the present work the discussion has been confined to bulk structures. The pn-junction device in this case is made of p- and n-material legs of macroscopic dimensions. Such structures have been demonstrated to convert heat into electricity over long periods of time.[10,11]

## 2.2) <u>Materials requirement evaluation criteria</u>:

The amount of materials required to convert a quantum of waste heat energy into useful electrical power can be effectively determined using the maximum power generation equation (4), without any reference to the efficiency of conversion, i.e. assuming that conversion takes place without any losses. Hence the minimum number of couples or pairs of pn-junction legs required to convert this quantum of energy can be determined using the rearranged relation;

$$N_{min} = \frac{4P_{max}^{N}R_{TED}}{\left(\alpha_{pn}\Delta T\right)^2} \quad (6).$$

It is seen from the above equation that the number of pn-junctions required to convert a quantum of heat into power depends on the device resistance which in turn depends on the junction legs geometry. The optimization of thermoelectric generator design therefore which includes the leg length, cross-sectional area of the legs, shape of legs and number of legs is non-trivial and complex.[15] The cross-sectional area of the two legs p- and n- need not be identical and can vary. Also, the cross-sectional shape can vary from the simple circular to polygonal and the legs need not have a constant cross-sectional area between the hot and cold junctions and hence can be tapered. All these factors result in varying the net heat distribution and thus alter the amount of power converted and efficiency of conversion. The power output and efficiency depend on the above mentioned parameters in a non-trivial manner and in order to keep the problem trackable, in the present work the main criterion considered for thermoelectric generator has been power output with simple geometry of legs. The following assumptions have been made to determine the amount of materials required for power generation;

1. The contact resistivity $\rho_c$ has been assumed to be zero.
2. The ratio of length to cross-sectional area of the device legs is assumed to be 0.5 $cm^{-1}$, 1.0 $cm^{-1}$ and 1.5 $cm^{-1}$. These values essentially consider the cases of short-wide legs and long-narrow slender legs as two extreme cases of aspect ratio.
3. The shape of the legs has not been explicitly factored and also the cross-sectional area has been considered to be a constant through the legs length.

## 3) <u>Results</u>:

### 3.1) <u>Waste heat categorization and quantity</u>:

The determination of amount of waste heat available globally in different sectors – transportation, industrial, residential and commercial is non-trivial. The quantum of rejected energy available every year has been quantified and mapped comprehensively by LLNL in the form of an energy flow diagram for the USA.[16] Tracking this rejected energy as well as the total energy consumed by USA for the past decade, Figure 2, clearly shows that waste heat increases steadily over the period of time and is of the order of ‘EJ’. This data however is not available for the entire world and more importantly, categorization or classification of this waste heat data based on type and quantity is a complex task, not easily undertaken by any organization or group. Also, there is no recent data available in open literature which has accurately analyzed waste heat data and classified them for the world. In a relatively recent study, the global waste heat data corresponding to the year 2012 was analyzed in detail including categorization and classification.[2] Although this data is a decade old with the current waste heat energy being higher, the methodology is conceptually relevant and so this data has been used in the present work. The available waste heat has been categorized into 3 types depending on its intensity as low temperature LT - < 100 °C (< 373 K), medium temperature MT – 100 °C to 300 °C (373 < T < 573 K) and high temperature HT - > 300 °C (> 573 K). The contribution to these categories comes from essentially 4 sectors mentioned above. The maximum total global waste heat in the year 2012 was found to be 340.512 EJ which when categorized into 3 types becomes 214.523 EJ, 54.483 EJ and 71.508 EJ respectively. The waste heat energy which is actually amenable however for utilization translates to 155.592 EJ, 39.429 EJ and 50.696 EJ with nearly 94.795 EJ not being amenable for utilization, Figure 3(a). At first glance this appears as though it is completely available to be converted or recovered. This however is not correct as the recovery or conversion processes are limited by the Carnot efficiency factor given by;

$$\eta_C = \left(\frac{T_h - T_c}{T_h}\right) \qquad (7).$$

This implies that the low grade heat is less efficiently recoverable compared to high grade heat. Considering $T_c$ to be 300 K in all the 3 cases, $\eta_c$ has been determined to be 0.196, 0.366 and 0.476 for the 3 temperature ranges LT, MT and HT respectively. Rationalizing with the Carnot efficiency factor to determine the actual recoverable heat energy, exergy has been determined and it is found to be 30.496 EJ, 14.431 EJ and 24.131 EJ respectively of low, medium and high grades, Figure 3(b). This clearly shows that a large quantity of waste heat cannot be recovered, specially the low grade heat which has a low Carnot efficiency factor.

In the present work this exergy available/accessible for conversion has been treated as $P_{max}^N$ to determine the number of pairs of legs $N_{min}$ required in each of the three categories. A careful literature search was undertaken in order to identify potential candidate materials, both p- and n- type which have been reported to exhibit the highest zT in the 3 temperature ranges and is given in Tables S1-S3. From among these, two p-type and two n-type alloys have been chosen for each of the temperature ranges, Table 1.[17] This selection has been solely based on zT and no other criteria was used. The temperature dependence of $\alpha$ and $\rho$ have been integrated over the reported temperature range to obtain average values given in Table 1 and these have been used to determine the number of pn-junctions required.

**3.2) <u>Thermoelectric materials – Quantity</u>:**

The maximum number of pn-junctions $N_{min}$ that will be required to convert the waste heat exergy into useful electrical power have been determined for the 3 temperature regions based on the materials properties given in Table 1 and given in Table 2, Figures 4(a)-(c). The number of junctions in all the cases is > $10^{11}$ and increases with increasing device resistance as the aspect ratio l/A increases from 0.5 $cm^{-1}$ to 1.5 $cm^{-1}$. The $N_{min}$ for low grade heat, < 373 K however is an order of magnitude higher compared to medium and high grade heat, commensurate with the magnitude of exergy available in these temperature regions. The quantity of different alloys required has been determined based on $N_{min}$ for each of the temperature regions and leg aspect ratio and is given in Table 3. It is seen that the quantity of alloys required increases from ~ 28 Mt to 280 Mt for the low temperature exergy while it increases from ~ 3 Mt to 30 Mt for high temperature exergy conversion. These are extremely large quantities of alloys required for waste heat conversion and should be weighed together with actual elemental requirement, world production and resources.

**4) <u>Discussion</u>:**

The assumption of zero contact resistivity is not practical and cannot be realized. However, it allows determination of minimum number of pairs N required to convert the exergy available and hence amount of materials required. Similarly, the assumption of matching load resistance with that of the device facilitates estimation of minimum amount of thermoelectric material required, which is the best case scenario. In practice however, the materials requirement will be far higher than these estimates, specially if one considers the conversion efficiency also which has been implicitly assumed to be maximum possible. These assumptions however facilitate easy determination of the minimum amount of materials required without compromising on accuracy. A detailed estimation of materials required will probably result in variations of the order of at least 10 % to 15 % more than the quantities determined in this work.

The different elements required to synthesize these alloys have been determined based on the alloys chemical composition and is given in Table S5. It is seen that a variety of elements will be required to synthesize the different thermoelectric alloys. The most prominent elements among them belong to groups 14-16 and are referred to as 'chalcogen' elements. Irrespective of which alloy is chosen for p-type or n-type and for any of the temperature ranges, the most critical elements are Pb, Bi, Te, Sn, Sb and Se which will be required in 'Mt' quantity as shown in Table 4. Unfortunately, the natural abundance of these elements in earth's crust is extremely small and hence materials availability will pose a risk. Also, most of these elements are produced as byproducts of transition metals production such as Cu and not as primary metals. The materials supply chain risk can be gauged from the world production statistics of these elements. The total amount of Bi produced in the world in the year 2021 was 19000 t while the world production of Te and Se are a mere 580 t and 3000 t respectively.[18] Among the groups 14-16 elements that are of importance to thermoelectrics, Sb is produced in relatively large quantities, i.e. 110,000 t in the year 2021. These production figures amply illustrate that elements requirement is several orders of magnitude compared to current production and availability. These elements are used for various applications other than thermoelectric energy conversion which renders elements availability to be even more highly acute. An additional criterion to be aware of with respect to these elements is the 'supply chain risk' with the risk assessment done based on the Herfindahl-Hirschman Index for the different elements (given in Supplementary). This index for the elements Bi, Te, Se, Sb and Sn has been determined using the production data by different countries in the year 2021. It is found that this index for all these elements is >> 2500, Table 4, clearly indicating the potential supply chain risk as they are produced predominantly by China. These data clearly show the materials supply chain risk associated with these elements thar are critical for converting the waste heat exergy into usable electricity. Even if one were to assume that just 10 % of the available exergy is converted into electricity, still it runs the risks of elements supply. The data on world resources on most of these elements is also not clearly available which makes it impossible to state that world production can be increased to meet the requirement.

Another aspect of sustainability of thermoelectric exergy conversion technology is the amount of energy required to produce these elements and the associated greenhouse gas emissions. Both these parameters – embodied energy and emissions for the groups 14-16 elements, except for Si, are not easily available as these elements are secondary products. Hence in the present work the embodied energy and emissions corresponding to primary product elements such as Cu and Pb[19,20] have been used to estimate the energy and emissions due to production of elements given in Table 4. The exact amounts depend on the type of alloys chosen for each of the temperature ranges and the aspect ratio of legs in the pn-junction. The embodied energy and emissions range from 50 PJ to 10831 PJ and the corresponding $CO_2$ emissions are of the order of 5 Mt to 744 Mt, Figures 5(a) & (b). These are extremely large quantities which go against 'sustainability' of thermoelectric energy conversion technology implementation.

The materials requirement, supply chain risk, embodied energy and emissions however have to be rationalized over the life time of this technology which currently stands at ~ 30 Years. If the energy and emissions are distributed over this 30 years period, then this technology is sustainable and green in nature. This argument however does not completely justify, as all the costs in terms of energy and emissions are upfront and are associated with initial implementation itself. The main constraint in which case will be the materials availability and supply chain risk for the realization of waste heat exergy conversion into useful electrical energy.

## 5) <u>Conclusions</u>:

The energy efficiency of almost all the processes, whether industrial or domestic, is < 100 % which results in energy waste to the extent of ~ 60 %. A significant fraction of this waste energy is in the form of heat and several technologies have been developed historically to 'harvest' this 'waste heat'. One such technique for harvesting is thermoelectric energy conversion which has several advantages. This technique however has not attained large scale implementation in spite of its advantages and the reason attributed is 'lack of materials with high figure-of-merit'. The recent developments in materials development has shown that figure-of-merit can reach 2.0 and in some alloys it can even reach 4.0. In light of these developments, the present work was undertaken to assess the potential as well as real challenges to implementing waste heat conversion using thermoelectric conversion technology.

A comprehensive analysis of the current literature was undertaken to select alloys/compounds which exhibit the highest figure-of-merit. The physical properties of these alloys were then used to determine the minimum number of pn-junctions required. It is found that $10^{11}$ to $10^{13}$ junctions will be needed depending on the magnitude of exergy, which is highest for low temperature. This translates to 'millions of tons' of alloys and hence 'millions of tons' of elements required. Among the different elements that will be required, elements such as Bi, Te belonging to groups 14-16 are critical and will be required to harvest any of the exergies – low, medium or high temperature. The current world production of these elements clearly shows that there will be a severe shortage of these elements and hence an extremely high probability for materials supply chain disruptions. Even if one were to assume that only a fraction of available waste heat will be harvested, the availability of these elements will severely restrict usage of thermoelectric energy conversion technology. The embodied energy and green house gas emissions due to extraction of these critical elements are also very high, indicating without ambiguity that this technology of waste heat recovery is non-sustainable. It is also clear from the current analysis that, figure-of-merit zT will not really be a limiting factor for large scale application if the alloys developed continue to have elements from groups 14-16 of the periodic table. The materials development strategy therefore needs a paradigm change and should be based on;

- Earth abundant and environmentally least intensive elements.

Both these criteria are equally important which can be illustrated with the e.g. $Mg_2Si$. This alloy has promising thermoelectric properties and has been extensively investigated. The two elements – Mg and Si are earth abundant but their synthesis is environmentally very intensive.

**Acknowledgements:**

The authors wish to acknowledge Indian Institute of Technology Bombay for the provision of facilities and a post-doctoral fellowship to AKB.

**Conflict of Interest:** The authors declare that there is no conflict of interest or competing financial interests.

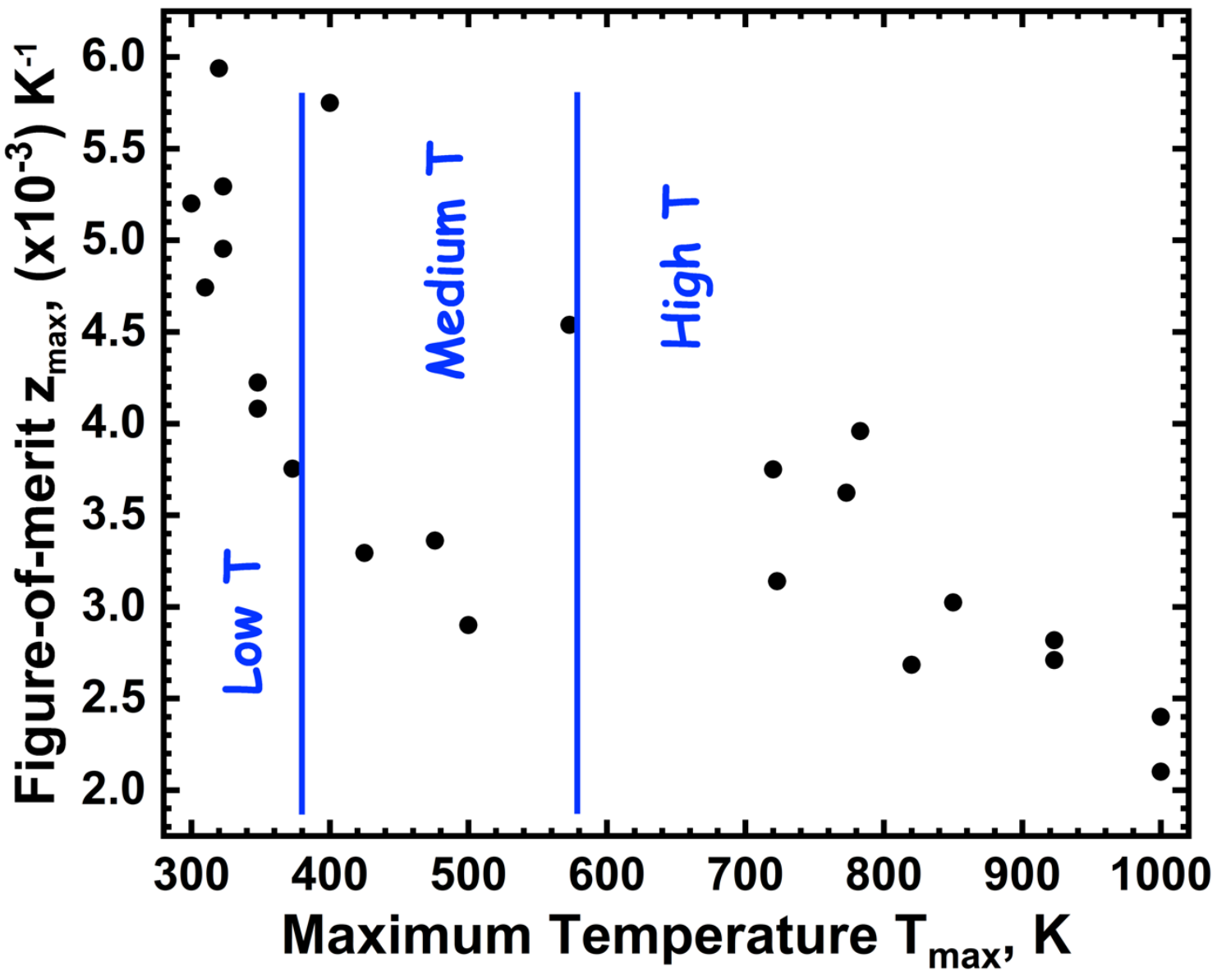

**Figure 1.** The thermoelectric figure of merit parameter $z_{max}$ analyzed for various temperatures shows clearly that several materials have been developed with technological potential for different temperatures.

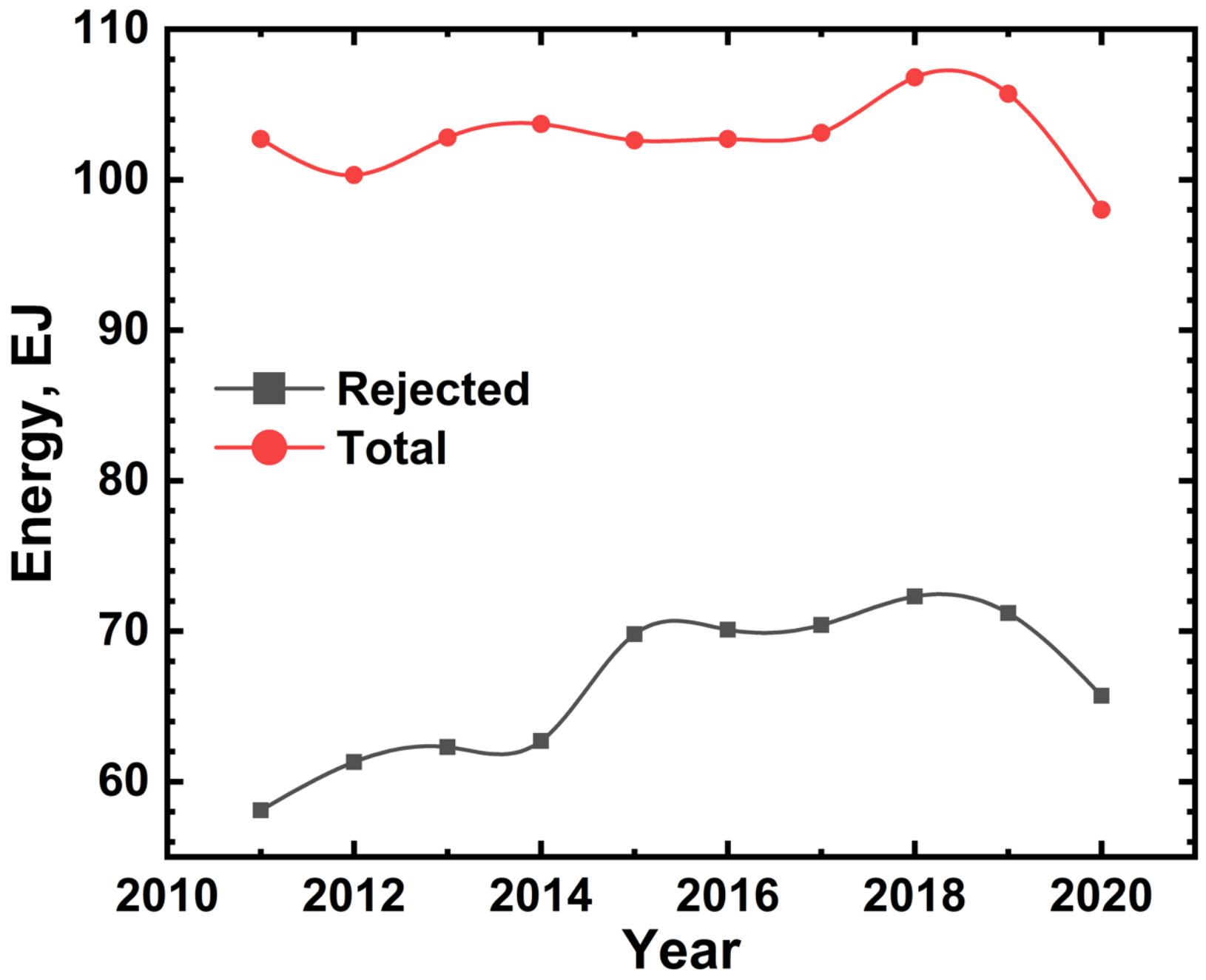

**Figure 2.** The total energy and the rejected energy by different sectors shows a steadily increasing behavior over the past decade for USA. This behavior is valid even for the world energy consumption.

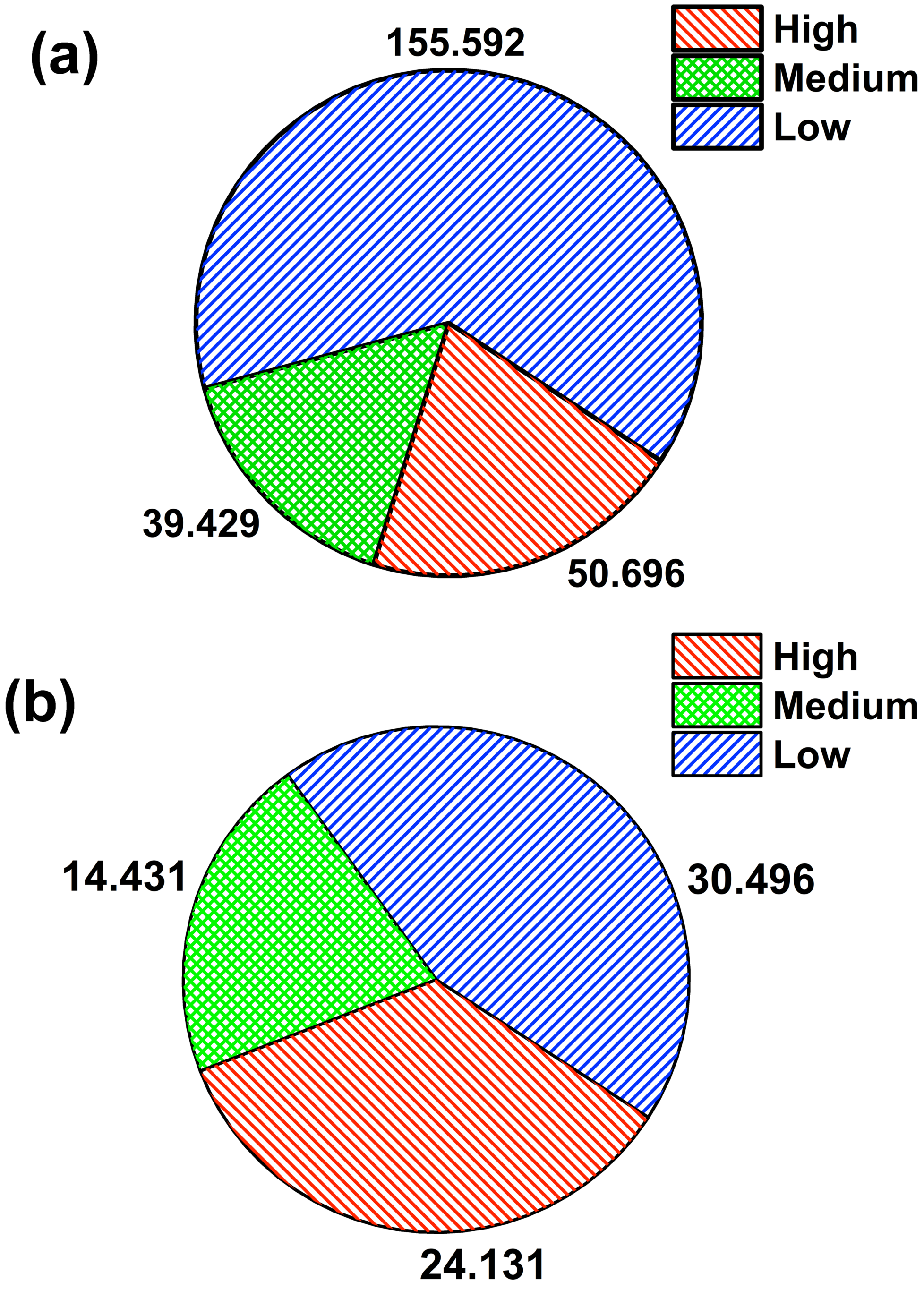

**Figure 3.** The amount of waste heat energy (EJ) amenable to recovery is highest at low temperature < 373 K while that available in the medium temperature range 373 K < T < 573 K is lowest, (a). The Carnot efficiency moderated energy, exergy (EJ) however is far less compared to the amenable energy (b).

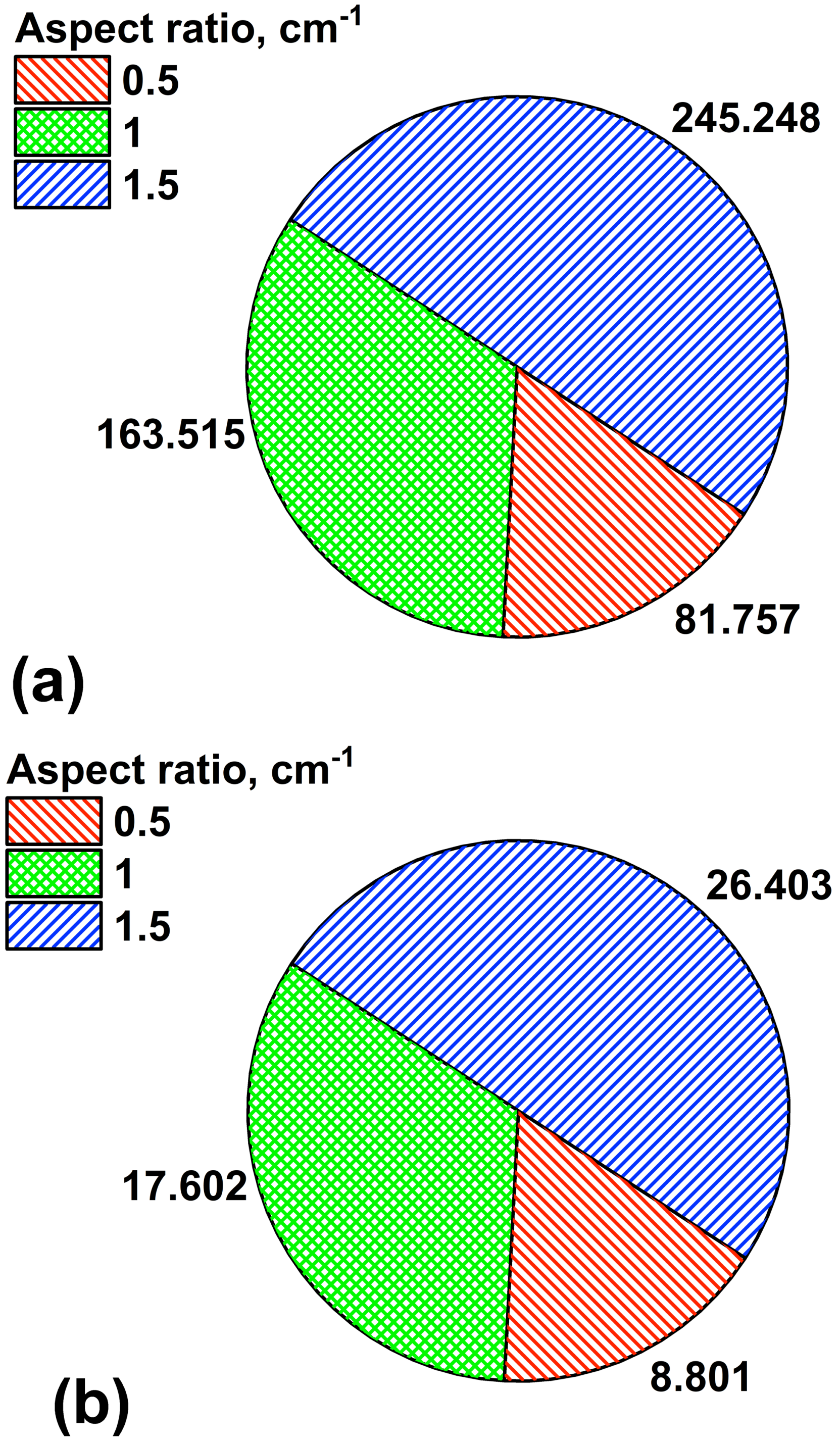

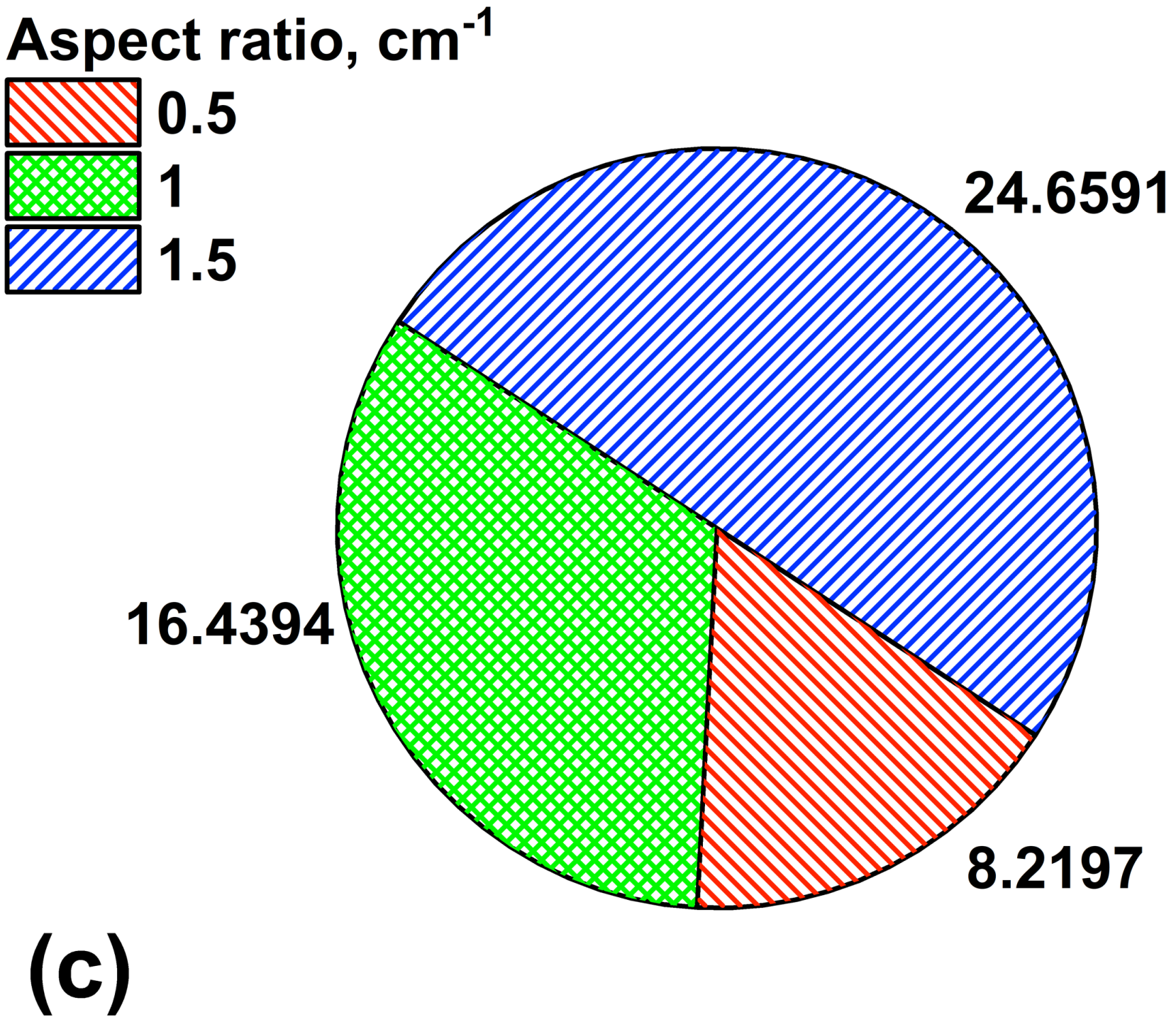

**Figure 4.** The minimum number of pn-junctions $N_{min}(x10^{11})$ required to convert exergy into electrical power for different aspect ratios l/A, $cm^{-1}$ and for different categories, (a) low temperature, (b) medium temperature and (c) high temperature. The number of junctions increases with increasing aspect ratio.

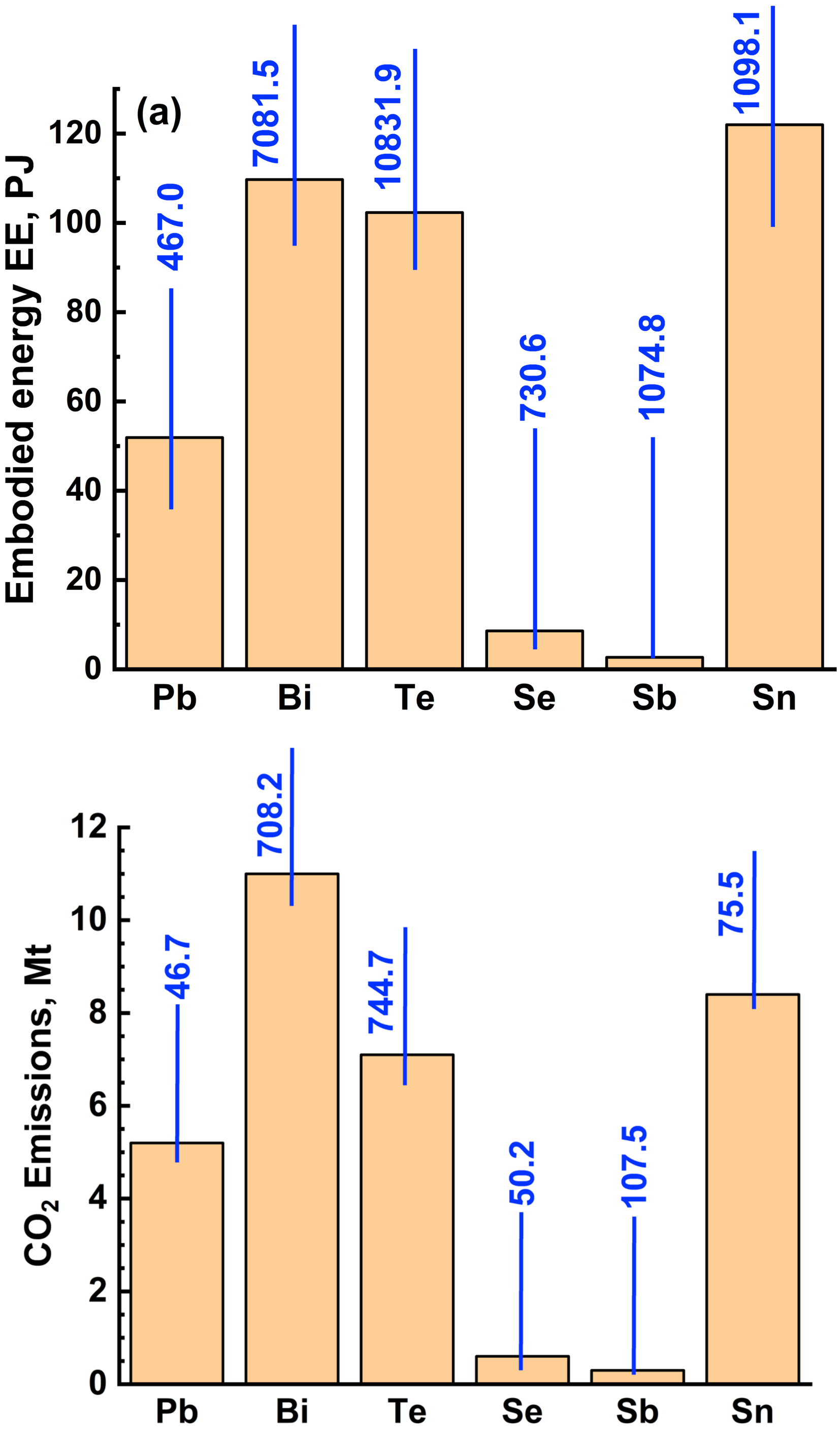

**Figure 5.** The embodied energy EE (a) and $CO_2$ emissions (b) due to extraction and manufacturing of different critical elements required for thermoelectric energy conversion. The EE and emissions given along vertical lines represent the upper limit depending on type of alloys selected for different temperature ranges.

**Table Captions:**

**Table 1.** Thermoelectric properties resistivity $\rho$, Seebeck coefficient $\alpha$ and figure-of-merit zT of select alloys which exhibit high zT at $T_{max}$ in low, medium and high temperature ranges.[17]

**Table 2.** The minimum number of pn-junctions $N_{min}$ required to convert the maximum exergy $P_{max}$ available in the three temperature ranges for 3 different aspect ratios l/A is given.

**Table 3.** The minimum quantity of different alloys/compounds required corresponding to the minimum number of junctions, pn-junctions $N_{min}$ shows that 'millions of tons' of alloys are needed to harvest the exergy. The quantity increases with increasing aspect ratio of the legs.

**Table 4.** The quantity of pure elements belonging to groups 14-16 which are critical for thermoelectric alloys synthesis are given along with their embodied energy and emissions due to manufacturing. The Herfindahl-Hirschman Index HHI indicates the material supply chain risk.

**Table 1:**

| Temperature, K | Alloy | ρ, x10$^{-5}$ Ωm | α, x10$^{-4}$ VK$^{-1}$ | $T_{max}$ | $(zT)_{max}$ |
|---|---|---|---|---|---|
| < 373 | $(Bi_{0.5}Sb_{1.5})Te_3$ (p) | 2.1003 | 2.3819 | 320 | 1.90 |
| | Mn-$Bi_2Te_3$ (p) | 2.5719 | 2.4094 | 310 | 1.47 |
| | $(BiSb)_2Te_3$ (n) | 0.6349 | 0.9032 | 303 | 1.28 |
| | $Bi_2(Te_{2.79}Se_{0.21})$ (n) | 1.2666 | 1.9451 | 357 | 1.20 |
| 373 < T < 573 | I-$Cu_2Se$ (p) | 1.7158 | 1.1595 | 400 | 2.30 |
| | $(AgSb_{0.94}Cd_{0.06})Te_2$ (p) | 4.3629 | 2.6676 | 573 | 2.60 |
| | $(Bi_{1.8}Sb_{0.2})(Te_{2.7}Se_{0.3})$ + 15 wt.% Te (n) | 1.2299 | 1.9383 | 425 | 1.40 |
| > 573 | $(NaEuSn)_{0.08}Pb_{0.92}Te$ (p) | 2.4928 | 2.1333 | 850 | 2.57 |
| | $(Ge_{0.87}Pb_{0.13})Te$ (p) | 0.6974 | 1.5272 | 723 | 2.27 |
| | $(SmMm)_{0.15}Co_4Sb_{12}$ (n) | 0.9873 | 1.8920 | 850 | 2.10 |
| | B-SnSe (n) | 0.1097 | 3.1476 | 773 | 2.80 |

**Table 2:**

| Temperature region | $N_{min}$, x$10^{11}$<br>Aspect ratio of Legs, l/A $cm^{-1}$ | | |
|---|---|---|---|
| | 0.5 | 1.0 | 1.5 |
| > 573 K | 8.2197 | 16.4394 | 24.6591 |
| 373 K – 573 K | 8.8010 | 17.6020 | 26.4030 |
| < 373 K | 81.757 | 163.515 | 245.248 |

Table 3:

| Temperature, K | Alloy | Quantity, Mt Aspect ratio of legs l/A, $cm^{-1}$ 0.5 | 1.0 | 1.5 |
|---|---|---|---|---|
| < 373 | $(Bi_{0.5}Sb_{1.5})Te_3$ (p) | 28.1244 | 112.4983 | 253.0956 |
| | Mn-$Bi_2Te_3$ (p) | 31.4765 | 125.9066 | 283.2614 |
| | $(BiSb)_2Te_3$ (n) | 28.0018 | 112.0078 | 251.9923 |
| | $Bi_2(Te_{2.79}Se_{0.21})$ (n) | 31.4765 | 125.9066 | 283.2614 |
| 373 < T < 573 | I-$Cu_2Se$ (p) | 3.0099 | 12.0398 | 27.0895 |
| | $(AgSb_{0.94}Cd_{0.06})Te_2$ (p) | 3.0892 | 12.3566 | 27.8024 |
| | $(Bi_{1.8}Sb_{0.2})(Te_{2.7}Se_{0.3})$ + 15 wt.% Te (n) | 3.3884 | 13.5535 | 30.4955 |
| > 573 | $(NaEuSn)_{0.08}Pb_{0.92}Te$ (p) | 3.3536 | 13.4146 | 30.1827 |
| | $(Ge_{0.87}Pb_{0.13})Te$ (p) | 2.6315 | 10.5262 | 23.6838 |
| | $(SmMm)_{0.15}Co_4Sb_{12}$ (n) | 3.0577 | 12.2309 | 27.5196 |
| | Br-SnSe (n) | 2.5399 | 10.1596 | 22.8589 |

**Table 4:**

| Element | Quantity, Mt | Embodied Energy, PJ | $CO_2$ Emissions, Mt | HHI |
|---|---|---|---|---|
| Pb | 2.08 – 18.68 | 51.9 – 467.0 | 5.2 – 46.7 | - |
| Bi | 4.39 – 283.26 | 109.7 – 7081.5 | 11.0 – 708.2 | 7158 |
| Te | 1.28 – 135.39 | 102.3 – 10831.9 | 7.1 – 744.7 | 3858 |
| Se | 0.11 – 9.13 | 8.6 – 730.6 | 0.6 – 50.2 | 2242 |
| Sb | 0.11 – 42.99 | 2.7 – 1074.8 | 0.3 – 107.5 | 3653 |
| Sn | 1.53 – 13.73 | 122.0 – 1098.1 | 8.4 – 75.5 | 1802 |

**Sustainability of largescale waste heat harvesting using thermoelectrics**

Anilkumar Bohra and Satish Vitta*
Department of Metallurgical Engineering and Materials Science
Indian Institute of Technology Bombay
Mumbai 400076; India.
* Email: satish.vitta@iitb.ac.in

## Supplementary Section

The recent developments in thermoelectric materials with high figure-of-merit zT has prompted an analysis of the potential of this technology for waste heat harvesting. To aid this analysis requires an extensive search of materials that exhibit high zT and hence the search was undertaken, the results of which are given in Tables S1, S2 and S3.[1] The materials have been categorised into low temperature < 373 K, medium temperature 373 K < T < 573 K and high temperature > 573 K in accordance with categorisation of waste heat. From among these alloys, two pairs from each of the category were selected for an in depth analysis of materials requirement and sustainability.

**Table S1**: List of alloys/compounds that exhibit high zT at low temperatures, < 373 K.

| Type | Alloy/Compound | $(zT)_{max}$ | $T_{max}$, K |
|---|---|---|---|
| p-type | $(AgCu)_{0.995}Te_{0.9}Se_{0.1}$ | 1.10 | 350 |
| | $(Bi_{0.26}Sb_{0.74})_2Te_3$+3%Te ingot | 1.12 | 298 |
| | BiSbTe | 1.20 | 300 |
| | $Bi_{0.4}Sb_{1.6}Te$ | 0.90 | 300 |
| | $(Bi_{0.2}Sb_{0.8})_2Te_3$ with 0.12 wt% multiwall carbon nanotubes | 1.47 | 348 |
| | $Bi_{0.5}Sb_{1.5}Te_3$ | 1.90 | 320 |
| | $Bi_{0.52}Sb_{1.48}Te_3$ | 1.56 | 300 |
| | $Bi_{0.5}Sb_{1.5}Te_3$ | 1.40 | 300 |
| | BiSbTe+0.75 wt% ZnAlO | 1.33 | 370 |
| | $Bi_{0.5}Sb_{1.5}Te_3$ | 1.20 | 363 |
| | $Bi_{0.5}Sb_{1.5}Te_3$ | 1.20 | 320 |
| | $Bi_{0.5}Sb_{1.5}Te_3$ | 1.42 | 348 |
| | $Bi_{0.5}Sb_{1.5}Te_3$+20 wt% Te | 1.24 | 350 |
| | $Bi_{0.4}Sb_{1.6}Te_3$+0.2 at% Te | 1.38 | 323 |
| | Mn doped $Bi_2Te_3$ | 1.47 | 310 |
| | $Bi_{0.4}Sb_{1.6}Te_3$ doped with 0.5 at% In | 1.20 | 320 |
| | $Bi_{0.48}Sb_{1.52}Te_3$ +$WSe_2$ | 1.27 | 360 |
| | $Bi_{0.5}Sb_{1.5}Te_3$ | 1.60 | 323 |
| | $Bi_{0.5}Sb_{1.5}Te_3$ | 1.00 | 300 |
| | $Bi_{0.48}Sb_{1.52}Te_3$ | 1.10 | 350 |
| | Polycrystalline $Bi_{0.5}Sb_{1.5}Te_3$ | 1.71 | 323 |
| n-type | Ternary $(BiSb)_2Te_3$ Bulk nanostructured | 1.28 | 303 |
| | $Ag_{0.011}Bi_2Te_{2.7}Se_{0.3}$ | 1.10 | 350 |
| | $Bi_2Te_{2.7}Se_{0.3}$ | 0.76 | Room Temperature |
| | $Bi_2Te_{2.7}Se_{0.3}$ | 0.94 | 300 |
| | $Mg_{3.2}Bi_{1.498}Sb_{0.5}Te_{0.002}$ | 0.75 | Room Temperature |
| | $Mg_{3.2}Bi_{1.498}Sb_{0.5}Te_{0.002}$ | 0.90 | 350 |
| | $Bi_2Te_{2.7}Se_{0.3}$ with0.8 atom % excess Bi | 0.88 | 348 |
| | $Bi_2Te_{2.79}Se_{0.21}$ | 1.20 | 357 |
| | $Bi_2Te_{2.85}Se_{0.15}$+0.08 wt% I | 1.07 | 360 |
| | $Bi_2Se_3$ | 0.96 | 370 |
| | $Ag_2S_{0.4}Se_{0.6}$ | 1.08 | 350 |
| | $Ag_2Se_{1.01}$ | 1.00 | 320 |

**Table S2:** List of alloys/compounds that exhibit high zT at medium temperatures, 373 K < T < 573 K.

| Type | Alloys/Compounds | $(zT)_{max}$ | $T_{max}$, K |
|---|---|---|---|
| p-type | $Cu_2Se$ | 0.72 | 380 |
| | $Bi_xSb_{2-x}Te_3$ | 1.40 | 373 |
| | $Bi_{0.3}Sb_{1.7}Te_3$+0.4 vol% SiC | 1.33 | 373 |
| | $Bi_{0.3}Sb_{1.7}Te_3$ | 1.30 | 380 |
| | $Bi_{0.5}Sb_{1.5}Te_3$ + Graphene | 1.05 | 425 |
| | $Bi_{0.4}Sb_{1.6}Te_3$+ 5wt% $Cu_7Te_4$ | 1.14 | 444 |
| | $Bi_{0.5}Sb_{1.5}Te_3$ +1 vol% $Cu_3SbSe_4$ | 1.60 | 476 |
| | $Bi_{0.4}Sb_{1.6}Te_3$ + $Cu_2Se$ nanocomposite | 1.60 | 488 |
| | $Cu_{1.97}Ag_{0.03}Se$ | 1.00 | 400 |
| | Iodine-doped $Cu_2Se$ | 2.30 | 400 |
| | MgAgSb | 1.00 | 423 |
| | $Mg_{0.97}Zn_{0.03}Ag_{0.9}Sb_{0.95}$ | 1.40 | 423 |
| | $MgAg_{0.965}Ni_{0.005}Sb_{0.99}$ | 1.40 | 450 |
| | $Mg_{0.995}Yb_{0.005}Ag_{0.97}Sb_{0.99}$ | 1.40 | 550 |
| | $AgSb_{0.94}Cd_{0.06}Te_2$ | 2.60 | 573 |
| | $MgAg_{0.97}Sb_{0.99}$ | 1.30 | 533 |
| | $Bi_2Te_3$ | 1.19 | 420 |
| n-type | $Bi_2Te_{2.7}Se_{0.3}$ | 1.04 | 399 |
| | $Cu_{0.01}Bi_2Te_{2.7}Se_{0.3}$ | 1.06 | 398 |
| | $Bi_2Te_{2.3}Se_{0.7}$ | 1.20 | 445 |
| | $Bi_2Te_{2.2}Se_{0.8}$ | 1.11 | 475 |
| | $Bi_2Te_{2.4}Se_{0.6}$ | 1.22 | 477 |
| | $Bi_{1.95}Sb_{0.05}Te_{2.3}Se_{0.7}$ | 1.30 | 450 |
| | $Bi_2Te_{2.7}Se_{0.3}$+16 wt% Te | 1.10 | 400 |
| | $Bi_{1.8}Sb_{0.2}Te_{2.7}Se_{0.3}$ + 15 wt% Te | 1.40 | 425 |
| | Lu doped $Lu_xBi_{2-x}Te_{2.7}Se_{0.3}$ | 1.37 | 373 |
| | $Mg_{3.3}Bi_{1.498}Sb_{0.5}Te_{0.002}$ | 0.90 | 423 |
| | $Mg_{3.3}Bi_{1.298}Sb_{0.7}Te_{0.002}$ | 1.20 | 548 |
| | $Mg_{3.2}Bi_{1.29}Sb_{0.7}Te_{0.01}$ | 1.24 | 573 |
| | $Mg_{3.2}Bi_{1.4}Sb_{0.59}Se_{0.01}$ | 1.24 | 498 |
| | $Ag_2Se$ with 0.5 wt% carbon nanotubes | 0.97 | 375 |
| | $Ag_2Se$ | 0.90 | 390 |
| | $Ag_2Se$ | 1.21 | 389 |
| | $Ag_2Sb_{0.01}Te_{0.99}$ | 1.40 | 410 |
| | $Ba_8Ga_{15.8}Cu_{0.033}Sn_{30.17}$ | 1.45 | 500 |

**Table S3:** List alloys/compounds that exhibit high zT at high temperatures > 573 K.

| Type | Alloys/Compounds | $(zT)_{max}$ | $T_{max}$, K |
|---|---|---|---|
| p-type | $Pb_{0.98}Na_{0.02}Te$- 8%SrTe | 2.50 | 923 |
| | $PbTe_{0.8}Se_{0.2}$ with 8% MgTe | 2.20 | 820 |
| | $Pb_{0.935}Na_{0.025}Cd_{0.04}Se_{0.5}S_{0.25}Te_{0.25}$ | 2.00 | 900 |
| | $Pb_{0.98}Na_{0.02}Se$-2%HgSe | 1.70 | 970 |
| | $Pb_{0.975}Na_{0.025}S$-3%CdS | 1.30 | 923 |
| | $(GeTe)_{0.95}(Sb_2Te_3)_{0.05}$ | 2.70 | 720 |
| | $Na_{0.03}Eu_{0.03}Sn_{0.02}Pb_{0.92}Te$ | 2.57 | 850 |
| | $Ge_{0.87}Pb_{0.13}Te$ | 2.27 | 723 |
| | $Ge_{0.87}Pb_{0.13}Te$-5%$Bi_2Te_3$ | 2.10 | 690 |
| | $AgCuTe_{0.90}Se_{0.10}$ | 1.60 | 670 |
| | $Sn_{0.91}Mn_{0.14}Te(Cu_2Te)_{0.05}$ | 1.60 | 920 |
| | $CuInTe_{1.99}Sb_{0.01}$+1.0 wt% ZnO | 1.61 | 823 |
| | $Cu_{0.7}Ag_{0.3}Ga_{0.4}In_{0.6}Te_2$ | 1.64 | 873 |
| | $Cu_2S_{0.52}Te_{0.48}$ | 2.10 | 1000 |
| | $DD_{0.7}Fe_3CoSb_{12}$ | 1.30 | 800 |
| | $(Nb_{0.6}Ta_{0.4})_{0.8}Ti_{0.2}FeSb$ | 1.60 | 1200 |
| | $Zn_{3.8}In_{0.2}Sb_3$ | 1.80 | 698 |
| | $Bi_{0.94}Pb_{0.06}Cu_{0.99}Fe_{0.01}SeO$ | 1.50 | 873 |
| | $Cu_{1.97}S$ | 1.70 | 1000 |
| | Polycrystalline SnSe | 3.10 | 783 |
| | SnSe Single Crystal | 2.60 | 923 |
| | $Cu_2Se$ with 0.75 wt % carbon nano tubes | 2.40 | 1000 |
| | $Mn_{30.4}Re_6Si_{63.6}$ | 1.15 | 873 |
| | $Mn_{1.06}Te$-2% SnTe | 1.40 | 873 |
| | $Yb_{14}Mn_{0.2}Al_{0.2}Sb_{11}$ | 1.30 | 1223 |
| | SiGe-$YSi_2$ nanocomposite | 1.81 | 1100 |
| **n-type** | Bi doped PbTe/$Ag_2Te$ (15%) | 2.05 | 800 |
| | $PbTe_{0.998}I_{0.002}$–3%Sb | 1.80 | 773 |
| | PbTe-4%InS | 1.83 | 773 |
| | $PbSe_{0.998}Br_{0.002}$-2%$Cu_2Se$ | 1.80 | 723 |
| | PbS-4.4%Ag | 1.70 | 850 |
| | $Pr_{2.74}Te_4$ | 1.70 | 1200 |
| | $Ba_{0.08}La_{0.05}Yb_{0.04}Co_4Sb_{12}$ | 1.70 | 850 |
| | $(Sm,Mm)_{0.15}Co_4Sb_{12}$ | 2.10 | 850 |
| | $Ti_{0.5}Zr_{0.25}Hf_{0.25}NiSn$ + 1 wt% DA | 1.50 | 825 |
| | $Ti_{0.5}Zr_{0.25}Hf_{0.25}NiSn_{0.998}Sb_{0.002}$ | 1.50 | 700 |
| | $Mg_{3.2}Sb_{1.5}Bi_{0.49}Te_{0.01}Cu_{0.01}$ | 1.90 | 773 |
| | Bromine doped SnSe | 2.80 | 773 |
| | Bi and Cr co doped $Mg_2Si_{0.3}Sn_{0.7}$ | 1.70 | 673 |
| | $La_{3-x-y}Yb_yTe_4$ | 1.20 | 1273 |
| | SiGe | 1.84 | 1073 |

The number of pn-junctions required to convert the maximum exergy available in the 3 temperature regions has been determined using Eq. (6) given in the main text. This number of junctions has then been converted into alloys and elements required using the procedure given below. The calculations shown below are an example for one particular case and is applicable to all the different cases studied.

**Low temperature exergy conversion:**

The cold side temperature $T_c$ of the thermoelectric device in all the cases is considered to be 300 K and hence the temperature difference ΔT that drives the conversion is 73 K with a Carnot efficiency factor of 0.196. The maximum exergy available therefore is 8.4718 PWh which translates to 0.9671 TW of power $P_{max}^N$ every year. Assuming the contact resistance to be zero, the device resistance $R_{TED}$ for an aspect ratio l/A of 0.5 $cm^{-1}$ has been determined using Equation (5) of main text by considering an average value of 2.336 and 0.951 mΩ $cm^{-1}$ as resistivities for p-type and n-type alloys respectively. This has been found to be 1.6435 mΩ. The average Seebeck coefficient is 0.2396 and 0.1424 $mVK^{-1}$ for the p-type and n-type respectively and the corresponding device Seebeck coefficient is 0.382 $mVK^{-1}$. These parameters have been used in Equation (6) of the main text to determine $N_{min}$ and it is found to be 81.757x$10^{11}$. Considering 1 $cm^2$ to be the cross sectional area of the leg results in the leg volume to be 0.5 $cm^3$ for an aspect ratio of 0.5 $cm^{-1}$. The density of $Bi_2Te_3$ is 7.7 $gcm^{-3}$ and hence each leg will weigh 3.85 g. The total amount of $Bi_2Te_3$ required to make 81.757x$10^{11}$ legs will be 31.4765 Mt. Therefore, the amount of elemental Bi and Te required will be 16.4307 and 15.0457 Mt corresponding to 52.195 wt.% and 47.8 wt.% in $Bi_2Te_3$ respectively.

**Table S4:** The minimum number of junctions required $N_{min}$ and amount of alloys and elements required to convert the exergy into electrical power is given. These are determined as per the procedure mentioned above.

| For low temperatures < 373 K | | | |
|---|---|---|---|
| | l/A=0.5 $cm^{-1}$ | l/A=1.0 $cm^{-1}$ | l/A=1.5 $cm^{-1}$ |
| $N_{min}$ | 81.757×$10^{11}$ | 163.515×$10^{11}$ | 245.248×$10^{11}$ |
| ***$Bi_2Te_3$ (Mt)*** | 31.4765 | 125.9066 | 283.2614 |
| **Bi (Mt)** | 16.4307 | 65.7232 | 147.8625 |
| **Te (Mt)** | 15.0457 | 60.1833 | 135.3990 |
| ***$(Bi_{0.5}Sb_{1.5})Te_3$ (Mt)*** | 28.1244 | 112.4983 | 253.0956 |
| **Bi (Mt)** | 4.3874 | 17.5497 | 39.4830 |
| **Sb (Mt)** | 7.6667 | 30.6670 | 68.9940 |
| **Te (Mt)** | 16.0703 | 64.2815 | 144.6190 |
| ***$(BiSb)_2Te_3$ (Mt)*** | 28.0018 | 112.0078 | 251.9923 |
| **Bi (Mt)** | 8.2018 | 32.8071 | 73.8086 |
| **Sb (Mt)** | 4.7771 | 19.1085 | 42.9899 |
| **Te (Mt)** | 15.0230 | 60.0922 | 135.1939 |
| ***$Bi_2(Te_{2.79}Se_{0.21})$ (Mt)*** | 31.4765 | 125.9066 | 283.2614 |
| **Bi** | 16.6416 | 66.5668 | 149.7603 |
| **Te** | 14.1738 | 56.6957 | 127.5526 |
| **Se** | 0.6610 | 2.6440 | 5.9485 |
| **For medium temperatures, 373 K < T < 573 K.** | | | |
| | l/A=0.5 $cm^{-1}$ | l/A=1.0 $cm^{-1}$ | l/A=1.5 $cm^{-1}$ |
| $N_{min}$ | 8.8010×$10^{11}$ | 17.6020×$10^{11}$ | 26.4030×$10^{11}$ |
| ***$Cu_2Se$ (Mt)*** | 3.0099 | 12.0398 | 27.0895 |
| **Cu (Mt)** | 1.8565 | 7.4261 | 16.7088 |
| **Se (Mt)** | 1.1534 | 4.6136 | 10.3807 |
| ***$(AgSb_{0.94}Cd_{0.06})Te_2$ (Mt)*** | 3.0892 | 12.3566 | 27.8024 |
| **Ag (Mt)** | 0.688 | 2.7518 | 6.1916 |
| **Sb (Mt)** | 0.730 | 2.9199 | 6.5697 |
| **Cd (Mt)** | 0.0429 | 0.1717 | 0.3864 |
| **Te (Mt)** | 1.6283 | 6.5132 | 14.6546 |
| ***$(Bi_{1.8}Sb_{0.2})(Te_{2.7}Se_{0.3})$ (Mt)*** | 3.3884 | 13.5535 | 30.4955 |
| **Bi (Mt)** | 1.6579 | 6.6318 | 14.9214 |
| **Sb (Mt)** | 0.1073 | 0.4294 | 0.9661 |
| **Te (Mt)** | 1.5187 | 6.0747 | 13.6681 |
| **Se (Mt)** | 0.1044 | 0.4177 | 0.9399 |
| **For High temperatures, > 573 K** | | | |
| | l/A=0.5 $cm^{-1}$ | l/A=1.0 $cm^{-1}$ | l/A=1.5 $cm^{-1}$ |
| $N_{max}$ | 8.2197×$10^{11}$ | 16.4394×$10^{11}$ | 24.6591×$10^{11}$ |
| ***PbTe (Mt)*** | 3.3536 | 13.4146 | 30.1827 |
| **Pb (Mt)** | 2.0756 | 8.3023 | 18.6801 |
| **Te (Mt)** | 1.2781 | 5.1123 | 11.5026 |
| ***$(Ge_{0.87}Pb_{0.13})Te$ (Mt)*** | 2.6315 | 10.5262 | 23.6838 |
| **Ge (Mt)** | 0.7637 | 3.0547 | 6.8731 |
| **Pb (Mt)** | 0.3255 | 1.3021 | 2.9297 |
| ***SnSe (Mt)*** | 2.5399 | 10.1596 | 22.8589 |
| **Sn (Mt)** | 1.5252 | 6.1009 | 13.7268 |
| **Se (Mt)** | 1.0147 | 4.0588 | 9.1322 |
| ***$Co_4Sb_{12}$ (Mt)*** | 3.0577 | 12.2309 | 27.5196 |
| **Co (Mt)** | 0.4364 | 1.7454 | 3.9271 |
| **Sb (Mt)** | 2.6214 | 10.4856 | 23.5925 |

The embodied energy EE and $CO_2$ emissions for most of the groups 14-16 elements are not available and these are typically secondary products of extraction of primary metals such as Cu and Pb. Hence, to determine the EE and emissions due to extraction of the elements required for thermoelectric conversion, typical values of Cu and Pb: 80 $GJt^{-1}$ and 25 $GJt^{-1}$ and 5.5 $tt^{-1}$ and 2.5 $tt^{-1}$ respectively have been used.[2,3] Te, Se and Sn are assumed to have EE and emissions similar to Cu while Bi and Sb are assumed to be similar to Pb.

The Herfindahl-Hirschman Index HHI has been determined using the relation;

$$HHI = \sum_{i=1}^{N} (S_i)^2 \qquad \text{(S1)}$$

where $S_i$ is the percentage share of world production of i and N is the total number of producers in the world.[4,5] The world production statistics of different elements have been obtained from USGS data base.[6]